\documentclass{ws-mplb}
\usepackage{amsfonts,amssymb,amsmath}
\usepackage{graphicx}
\usepackage[super,sort,compress]{cite} 

\begin{document}

\markboth{Authors' Names}{Instructions for typing manuscripts (paper's title)}

%
\catchline{}{}{}{}{}
%

\title{Hamiltonian mean field model : effect of temporal perturbation in coupling matrix\footnote{For the title, try not to
use more than 3 lines. Typeset the title in 10 pt
Times Roman, boldface.}
}
\author{\footnotesize Nivedita Bhadra\footnote{Typeset names in
10~pt Times Roman. Use the footnote to indicate
the present or permanent address of the author.}}

\address{Department of Physical Sciences, Indian Institute of Science Education and Research Kolkata\\
Mohanpur, Nadia, West Bengal 741246, India\footnote{State completely without abbreviations, the
affiliation and mailing address, including country. Typeset in 8~pt
Times italic.}\\
nivedita.home@gmail.com}

\author{Soumen K Patra}

\address{Department of Physical Sciences, Indian Institute of Science Education and Research Kolkata\\
Mohanpur, Nadia, West Bengal 741246, India\\
soumen.k.p@gmail.com}

\maketitle

\begin{history}
\received{(Day Month Year)}
\revised{(Day Month Year)}
\end{history}

\begin{abstract}
The Hamiltonian mean-field model is a system of fully coupled rotators which exhibits a second-order phase transition at some critical energy in its canonical ensemble. 
We investigate the case where the interaction between the rotors is governed by a time-dependent coupling matrix. Our numerical study reveals a shift in the critical point due to the temporal modulation.The shift in the critical point is shown to be independent of the modulation frequency above some threshold value, whereas the impact of the amplitude of modulation is dominant. In the microcanonical ensemble, the system with constant coupling reaches a quasi-stationary state at an energy near the critical point. Our result indicates that the quasi-stationary state subsists in presence of such temporal modulation of the coupling parameter.
\end{abstract}
\section{Introduction}
\label{sec:intro}
Hamiltonian mean field model (HMF) represents a conservative system with long-range interactions of particles moving on a circle coupled by a repulsive
or attractive cosine potential \cite{Antoni1995,Campa2009,Campa2014,Levin2014}. This coupled rotator system exhibits several unusual properties, such as the
presence of quasi-stationary states (QSS) characterized by anomalous diffusion, vanishing Lyapunov exponents, non-Gaussian velocity distributions, aging and fractal-like phase space structure etc\cite{Latora2001}. The underlying physics of this coupled system explains several real-life phenomena in free electron lasers\cite{Antoniazzi2006}, rarefied plasmas\cite{Elskens2003,Benedetti2006}, beam particle dynamics, the gravitational many-body problem\cite{Lynden-Bell1967}. This system is an example of long-range interaction and shows a \emph{second-order phase transition} from a clustered phase to a gaseous one (where the particles are homogeneously distributed on a circle) as a function of energy. Prior to its equilibrium state, the system presents several interesting features. For example, if the particles are prepared in a ``water bag" initial state, the relaxation to the equilibrium becomes very slow\cite{Montemurro2003,Pluchino2004a,Rapisarda2005,Pluchino2005,Pluchino2006,Tauro2007,Antoniazzi2007,Leoncini2009}.
The relaxation time of this \emph{quasi-stationary states} diverges with the system size $N$.
In our present work, we introduce a temporal perturbation to the coupling and investigate the effect on the critical point and QSS. Our present work is focussed on the effect on the equilibrium phase transition. However, we have briefly discussed the out-of-equilibrium phase transition and report non-Gaussian velocity distribution in presence of the temporal modulation. Our numerical investigation qualitatively discusses the finite size effect on the relaxation time in this context.
\par
\hspace{1.5cm} In conventional HMF model, the moments of inertia of the rotors are usually considered to be time independent, identical, isotropic and they are equally interacting to all
other rotors\cite{Antoni1995,Barre2006,Campa2007,Konishi1992}. 
Such simplified model with uniform constant coupling provided several important insights about systems with long-range interaction. However, interactions are rarely uniform and heterogeneity in the coupling strength is more realistic in real-life for a system with long-range interaction e.g., stars and self-gravitating system has heterogeneous mass distribution, vortices in 2D turbulence have a heterogeneous circulation. The interaction is encoded in the coupling matrix and network topology of this matrix plays a crucial role in determining the critical behaviour of this thermodynamic system. It has been reported that the critical energy depends on the network parameter in such a system\cite{Medvedyeva2003,Kim2001,DeNigris2013}. Some recent studies have shown that spectral properties of coupling matrix plays significant role in the synchronization of this system\cite{Restrepo2014,Virkar2015}.
HMF model on Erd\"os-Renyi networks was studied in Ref.\cite{Ciani2010}. Nigris-Leoncini\cite{DeNigris2013} studied the model with Watt-Strogatz small-world network\cite{Watts1998}. Importance of link density  in a network to understand such a system with long-range interaction has been addressed in Ref. \cite{Ciani2010,Luo2011,Nigris2013}.\par 
A natural question is, what happens when the elements of the coupling matrix are time-dependent? How does it influence the second-order phase transition? What about the quasi-stationary state in presence of this imposed temporal modulation?
In order to address this issue, we consider a HMF model where each element of the coupling matrix is temporally and periodically modulated. Although a majority of works in such long-range system focus on static networks, there are applications where coupling strength and network topology can vary in time \cite{Stojanovski1997,Ito2001,Zanette2004,Masuda2016}. 
A recent interesting study by Petit et al.\cite{Petit2017} shows how tuning the topology of a time-varying network leads to the emergence of self-organised pattern in the system. Another research work\cite{Stilwell2006} discusses time-varying network topology in the context of synchronization.
In this paper, our focus is on the effect of the temporal modulation on the phase transition in HMF model and in its inequilibrium state.  
The coupling is taken in such a way that it is always positive in magnitude. For the HMF model with constant coupling, we observe a second-order phase transition in its canonical ensembles at the critical energy $U_c\approx0.75$. In contrast to that, it occurs at $U_c\approx 0.49$ for the temporally modulated case. We perform N-body numerical simulations to study the system. Our numerical analysis shows that the system is insensitive to the change of the modulation frequency above a certain threshold value, whereas, the effect of amplitude is dominant. Finite size effect, trapping probability have been discussed in this context. In addition to that, our numerical study of the dynamics of the microcanonical ensemble hints at the persistence of QSS state in presence of the imposed modulation. We report the presence of symmetric bump in the velocity profile and qualitatively discuss the finite size effect on the relaxation time in the QSS regime.  
\par
 The rest of the paper is organized as follows. In Sec. \ref{sec:model} we introduce the standard HMF model and HMF model with periodic modulation. We present our numerical results in Sec. \ref{sec:numeric}. 
In Sec.\ref{sec:concl} we have drawn conclusions from the numerical results.

\section{Model}
\label{sec:model}
The Hamiltonian describing the HMF model with constant coupling is given by
\begin{eqnarray}
\label{eq:HMFmodel}
H=\sum_{i=1}^N\frac{p_i^2}{2}+\frac{\epsilon}{2N}\sum_{i,j=1}^N(1-\cos(\phi_i-\phi_j))=K+V,
\end{eqnarray}
where $\phi_i\in[-\pi,\pi]$ is the angle that particle $i$ makes with a reference axis and $p_i$ stands for its conjugate momentum. This is a system of identical particles on a circle of unit mass or a classical XY rotor system with infinite range coupling.  For $\epsilon>0$ (attractive), the rotor tends to align (ferromagnetic case) whereas for $\epsilon<0$ (repulsive), spin tends to anti-align (anti-ferromagnetic case). The first term is the kinetic energy $K$ and the second term $V$ is the interaction energy which is scaled by the total number of particles $N$ to make it thermodynamically stable. The $1/N$ factor in the potential energy makes the energy  extensive (Kac prescription) and justify the validity of mean field approximation in the limit $N\rightarrow \infty$. \par 
To understand the physical meaning of this system, the standard method is to consider a  mean field vector $\mathbf{M}$
\begin{eqnarray}
\label{eq:MFvector}
\mathbf{M}=Me^{i\phi}=\frac{1}{N}\sum_{i=1}^N \mathbf{m}_i,
\end{eqnarray}
where, $\mathbf{m}_i=(\cos\phi_i,\sin\phi_i)$\cite{Antoni1995}. $M$ and $\phi$ are modulus and phase of the order parameter which specifies the clustering of particles or for XY model, it is the magnetization. The potential energy can be rewritten as a sum of single particle potentials $v_i$
\begin{eqnarray}
V=\frac{1}{2}\sum_{i=1}^N v_i,
\quad v_i=1-M\cos(\phi_i-\varphi). 
\end{eqnarray}
 This model is exactly solvable at equilibrium, where a second-order phase transition is observed. The transition is from a low energy condensed phase or ferromagnetic phase with $M\neq 0$, to a high energy phase or paramagnetic phase with $M=0$. The transition can be quantified from the caloric curve
 \begin{eqnarray}
 \label{eq:caloriccurve}
 U=\frac{\partial(\beta f)}{\partial \beta}=\frac{1}{2\beta}+\frac{\epsilon}{2}(1-M^2),
\end{eqnarray}  
where $\beta=1/K_B T$, $K_B$ being the Boltzmann constant. Considering $\epsilon=1, \beta=2$, a transition is found at $U=U_c=0.75$. 
At variance with this scenario, we observe several departures from this if we introduce a temporal modulation to the coupling. We consider the modulation of the form $\epsilon(t)=|a \cos{\omega t}|$. Hamiltonian governing this problem is given by
\begin{eqnarray}
\label{eq:HMFmodel_modulated}
H=\sum_{i=1}^N\frac{p_i^2}{2}+\frac{\epsilon(t)}{2N}\sum_{i,j=1}^N(1-\cos(\phi_i-\phi_j))=K+V(t).
\end{eqnarray}
 Assuming the similar mean filed vector, we write the Hamiltonian for this model as
\begin{eqnarray}
H(t)=\sum_{i=1}^N\frac{p_i^2}{2}+\epsilon(t) M\sum_{i=1}^N(1-\cos(\phi_i-\varphi)).
\end{eqnarray}
Therefore, the equations of motion for this system are
\begin{subequations}
\label{eq:modulatedHMFEOM}
\begin{align}
\dot{\phi}_i&=p_i,\\
\dot{p_i}&=-\epsilon(t)M\sum_{i=i}^N\cos(\phi_i-\varphi).
\end{align}
\end{subequations}
These equations are essentially Eq. of motion for fully coupled driven pendula. In the next section, we would integrate these equations to study the dynamics of the system.
\section{Numerical results}
\label{sec:numeric}
\begin{figure}
\centering
\includegraphics[scale=.5]{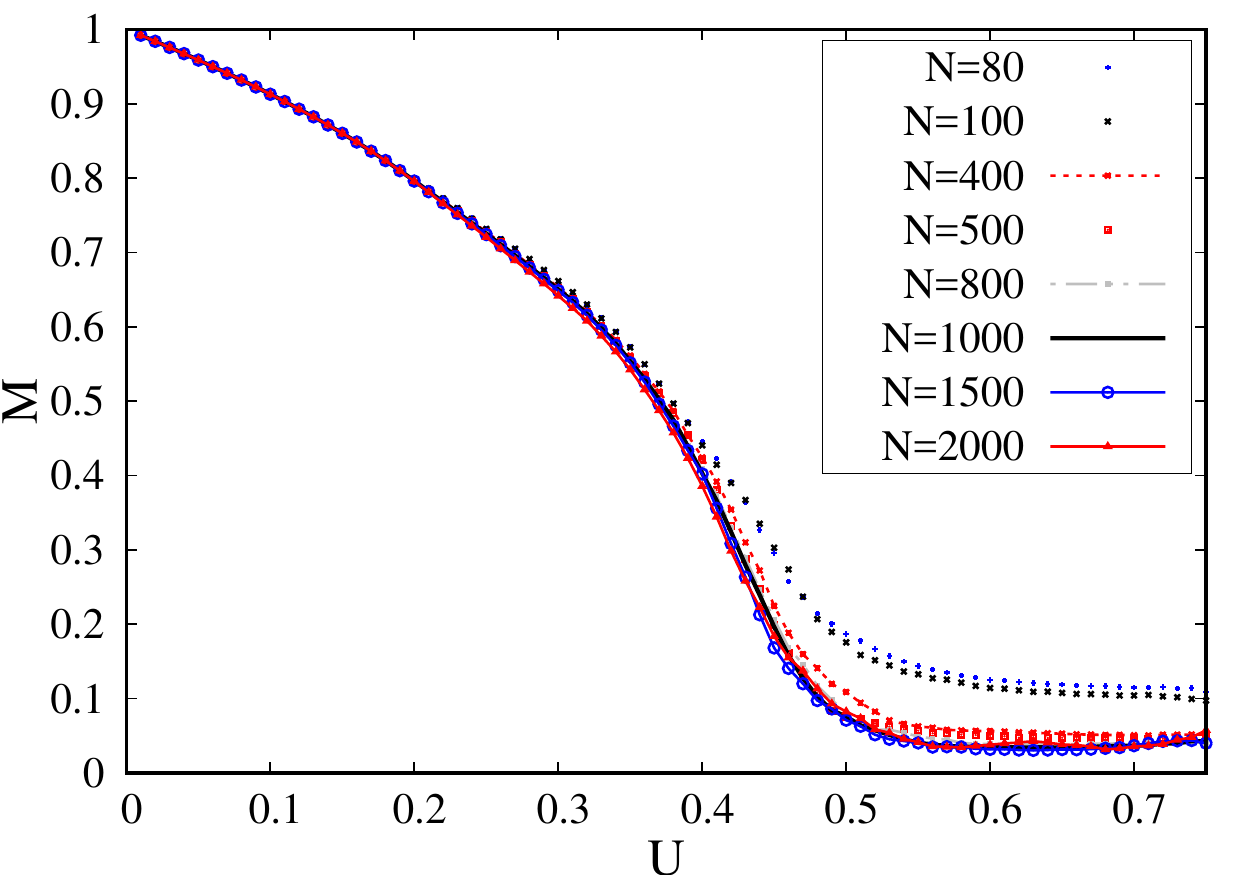}
\caption{Figure plots modulus of magnetization $M$ as a function of initial energy $U$ for $N = 80$ to $2000$. Symbols represents $M$ vs $U$ plot for different system size $N$. Each point is numerically simulated and averaged over value of $20$ different initial conditions. Parameters are $a = 1, \omega = 10$. }
\label{fig:fseffect}
\end{figure}

In order to study the effect of temporal modulation to the coupling parameter, we integrated the equation of motion (Eq. \ref{eq:modulatedHMFEOM}) numerically by adopting RK4 algorithm with a fixed time step $dt=0.01$ and total integration time $t_f = 1000$, which includes a period during which transients decay. Initial configuration was chosen as follows : positions of individual particles $\phi_i$ are uniformly distributed over the circle with a zero mean value and all the momenta $p_i$ are scaled in to attain the desired initial total energy $E$. To obtain the asymptotic behaviour, we computed the time-average for both magnetization $M$ and kinetic energy $T$ once the transient lapses. It should be mentioned that the transients depend on the system size $N$ and total initial energy $E$ which we have checked for individual cases. However, the typical integration steps to reach a good convergence was of the order of $10^5$ and averaging was performed for $10^3$ time steps to calculate $M$ and $T$.
\begin{figure*}[hb]
\includegraphics[width=6cm]{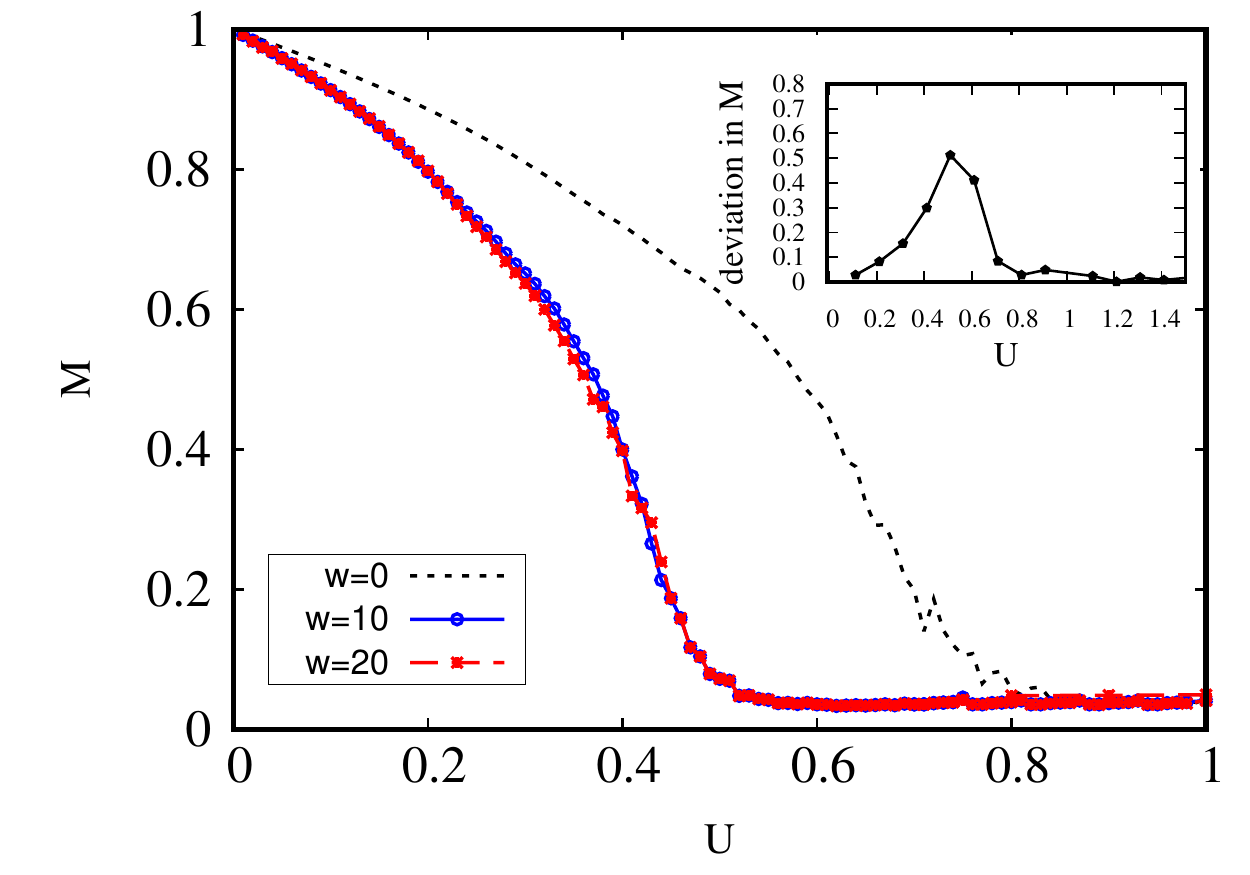}(a)\includegraphics[width=6cm]{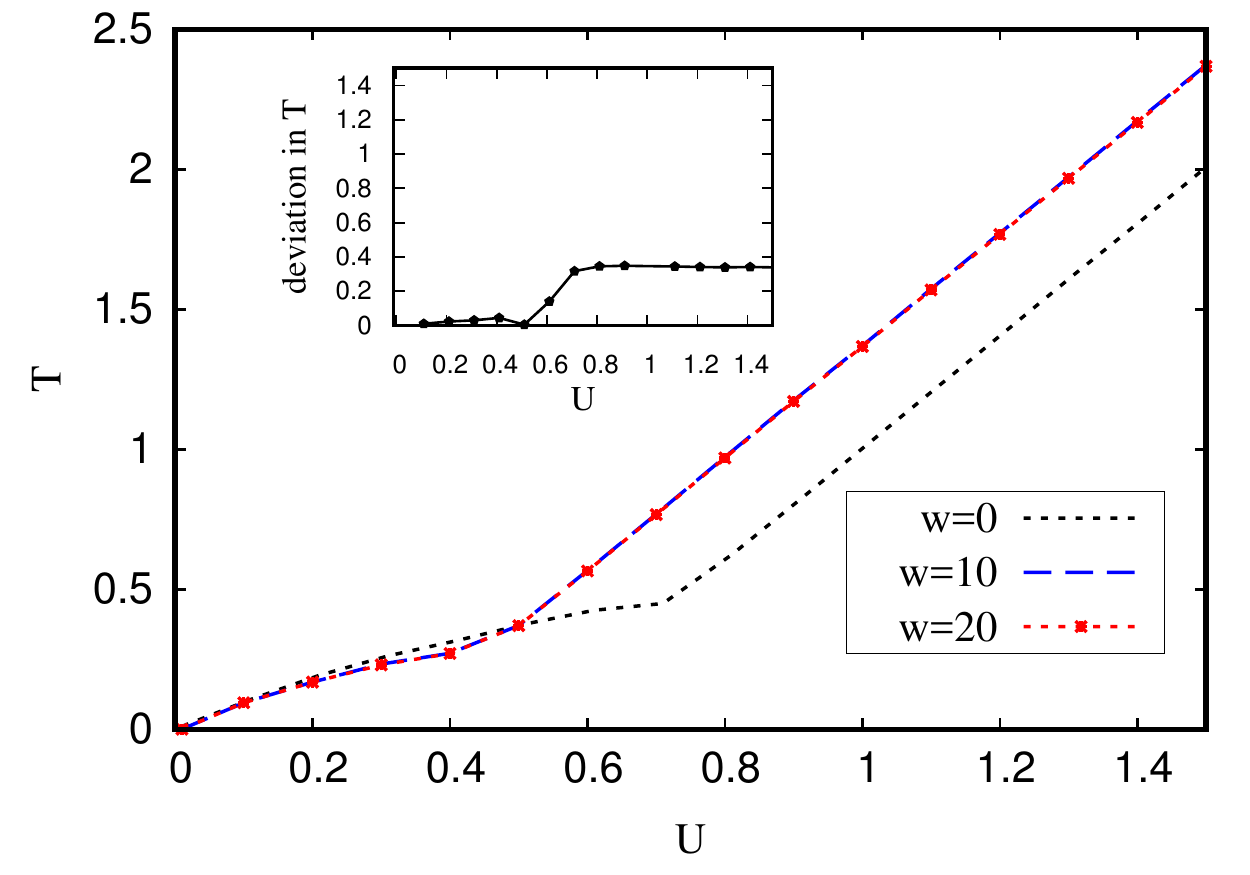}(b)
\vspace*{8pt}
\caption{ (a) Modulus of magnetization $M$ as a function of  initial energy $U=H/N$ for $N=1000$. Each point is numerically obtained after averaging over 20 different initial conditions.  Inset figure shows how the magnetization is deviated in the modulated case from the constant coupling case. (b) Variation of kinetic energy $T=2\langle K\rangle /N$ as a function of initial energy $U$ for $N=1000$. Each point corresponds to averaging over different initial conditions (typically $20$). Inset figure shows the deviation of  kinetic energy($T$) from the constant coupling case.
}
\label{fig:MavgvsU}
\end{figure*}
 Magnetization $M$ is calculated using Eq. \ref{eq:MFvector}. Here, we considered the system size $N=1000$ and kept $a=1$.
 In order to analyze the finite size effects, we vary the system size from $N=80$ to $N=2000$ when $\omega = 10$. Fig. \ref{fig:fseffect} shows variation of average magnetization vs energy per particle for different system size. For each system size, there is a order-disorder transition at $U_c\approx0.49\pm\delta$, where $\delta$ is $\mathcal{O}(10^{-3})$. However, increasing the system size $N$ the magnetization gradually falls down to zero for $U>U_c$.
 \par
 Fig.\ref{fig:MavgvsU} displays the absolute magnetization $M$ as a function of initial energy $U (E/N)$ per particle for both $\omega = 0$ and $\omega\neq 0$. Here, we have shown specifically for $\omega= 10$, and $20$. 
 We observe that in both cases the system undergoes a phase transition from a ferromagnetic/clustered phase with $M\neq 0$ to a paramagnetic/homogeneous phase reaching $M\approx 0$ at some critical values of initial energy $U$. It is known that for $\omega=0$, critical energy is $U_c = 0.75$ and critical temperature $T_c = 0.5$ separating the two phases \cite{konishi100}. Our numerical study shows $U_c \approx 0.76$ for $\omega = 0$ and for both $\omega =10$ and $\omega = 20 $, $U_c=0.49$. Critical energy $U_c$ is calculated by linear-curve fitting method.
\begin{figure}[ht]
\centering
\includegraphics[width=5.7cm]{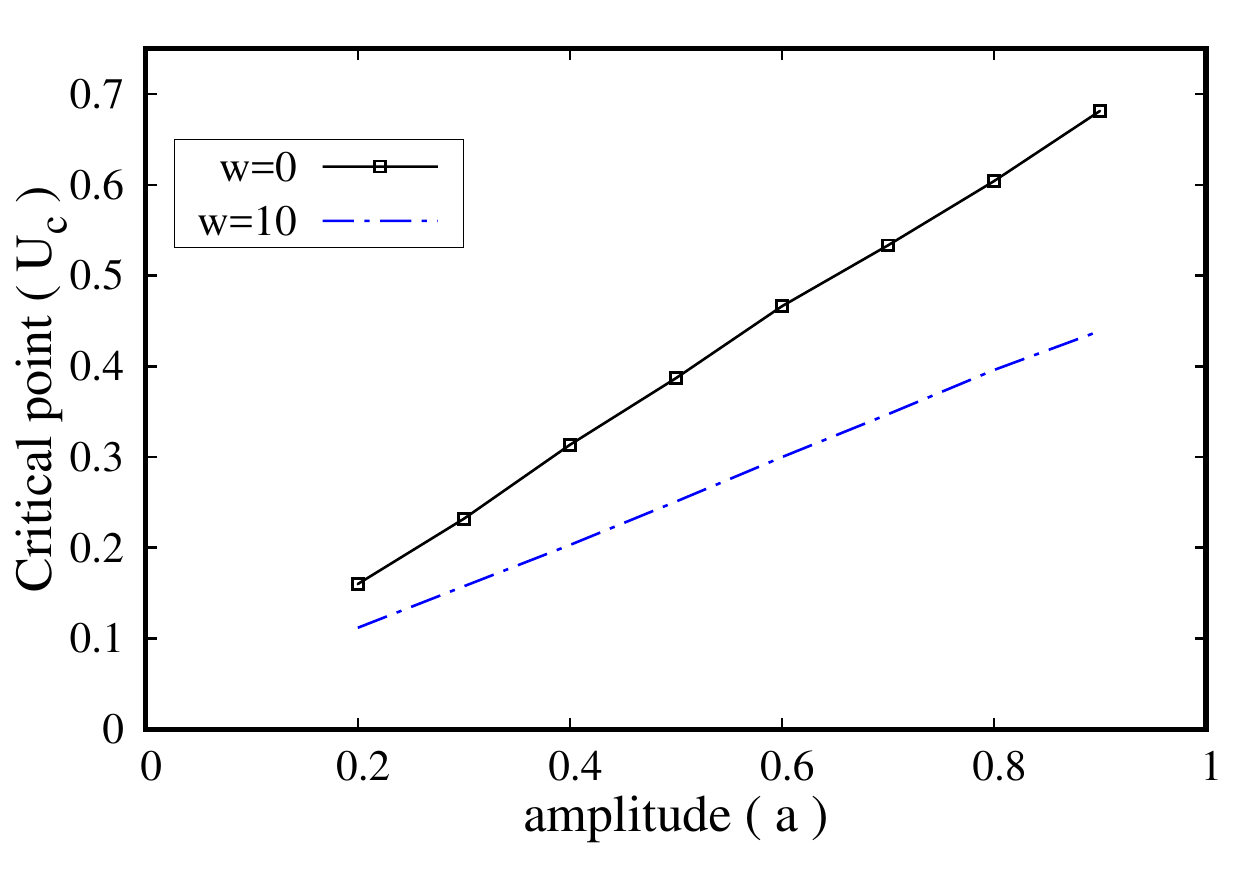}(a)
\includegraphics[width=5.7cm]{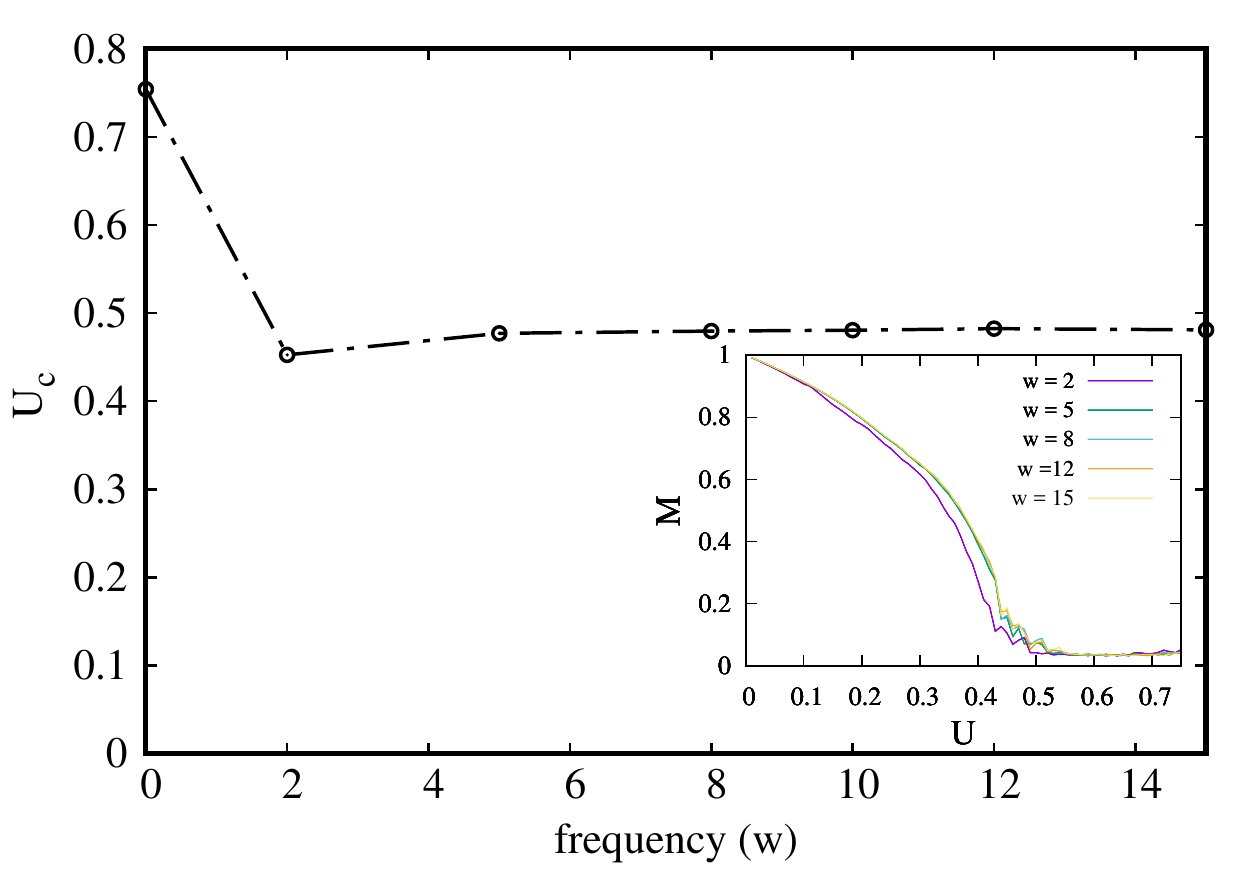}(b)
\vspace*{8pt}
\caption{ (a) Variation of critical point $U_c$ with amplitude of modulation $a$ for $\omega=0$ and $\omega=10$. Solid line corresponds to constant coupling case and dot-dashed line represents the variation for temporally modulated case.   (b) Figure shows variation of the critical point($U_c$) with frequency of modulation. Inset figure shows modulus of $M$ vs $U$ plots for different modulation frequencies ($\omega$). In both cases, system size $N = 10^3$. Averaging was done typically over $20$ different realizations. For each case, $U_c$ is obtained by linear curve-fitting method.
}
\label{fig:Ucscaling}
\end{figure} 
The deviation from the $\omega = 0 $ case in the $M$ vs $U$ curve is shown in inset figure of Fig. \ref{fig:MavgvsU}. Note that it is maximum around $U_c$. 
\begin{figure}[h]
\centerline{\includegraphics[width=7.5cm]{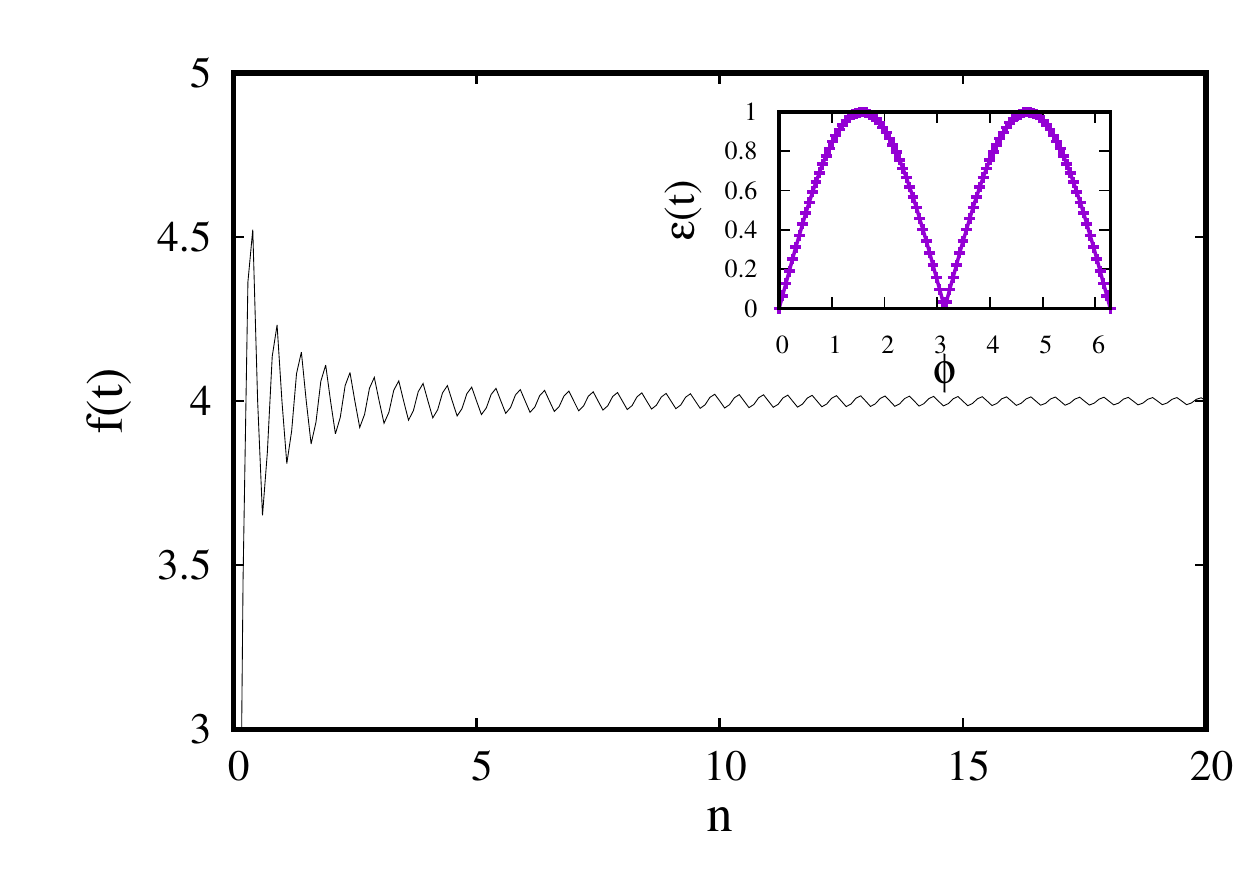}}
\vspace*{8pt}
\caption{Figure shows variation of $f(t)=\int |a \cos(n\omega t)| dt$ with $\omega$ for $ 0< \omega < 20$. Inset shows the function $\varepsilon(t)$ as a function of time($t$). The change in the value of function for $\omega>2$ is $10^{-2}$ which is negligible compared to the mean value of it.}
\label{fig:datafluctuation}       
\end{figure}
We vary amplitude $a$ and frequency of modulation $\omega$ and obtain the critical point $U_c$ and compare the results with $\omega=0$ case. The shift of the critical point $U_c$ for the system is found to be insensitive to $\omega$ above $\omega = 2$ which is shown in Fig. \ref{fig:Ucscaling}(b). This can be understood from the variation of the integrand $f(t)$ with the modulating frequency as shown in Fig.\ref{fig:datafluctuation}. The variation of the function $f(t)$ is negligible compared to the mean value of $f(t)$ when $\omega>2$. The effect of amplitude is, in contrast, dominant. We compare the variation of $U_c$ with the change in $a$ for both cases. In both cases, $U_c$ changes linearly with $a$ with different slopes as shown in Fig. \ref{fig:Ucscaling}(a).\par 
\begin{figure}[ht!]
\centerline{\includegraphics[width=6cm]{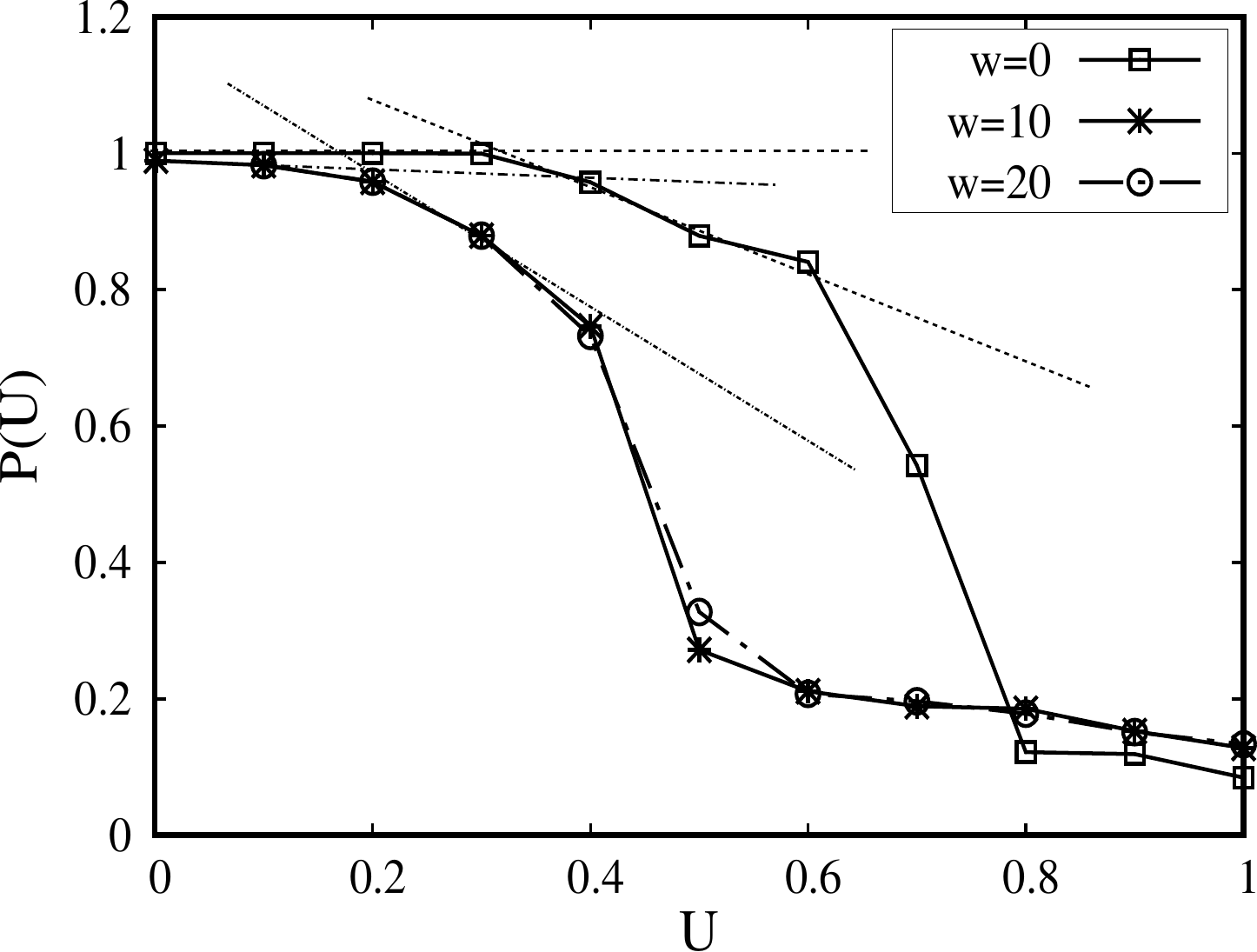}}
\vspace*{8pt}
\caption{Trapping probability $p(U)$ as a function of energy $U=H/N$. Symbols refer to $p(U)$ for $\omega= 0, 10$ and $20$. Each point represents numerically calculated $p(U)$ over typically $20$ different realizations. The horizontal straight segment is drawn to show  the regime $p(U)\approx 1$, where most of the particles are trapped.The other segment guides the eyes through decreasing values of $p(U)$.
}
\label{fig:trapping}
\end{figure}
 To explore more about the dynamics of the particles and distribution of energy among them, we calculated \emph{trapping probability} following Ref. \cite{Antoni1995}. This is an analogous quantity to \emph{activity parameter} \cite{Escande1994}. The particles are catagorised into (i) high energy particles (HEP) having energy $e_i>\epsilon (1+M)$ and (ii) low energy particles (LEP) having energy $e_i<\epsilon (1+M)$. The fraction of LEP i.e., $N_{\text{LEP}}/N$ is defined as the trapping probability once the system reaches equilibrium. $N_\text{LEP}$ are the particles which belong to the regime within the separatrix in the phase space of Eq. \ref{eq:modulatedHMFEOM}. In the phase space, the energy of the particles at the separatrix is $\epsilon(1+M)$. Particles having energy below this remains bounded within the separatrix (subcritical regime), whereas, higher energy particles can visit the whole circle (supercritical regime). Fig. \ref{fig:trapping} plots the energy dependence of trapping probability $p(U)$. We considered system size $N = 10000$ for estimating $p(U)$. The horizontal segment in the figure represents the energy regime when $p(U)\approx 1$, implying that majority of the particles are trapped in the subcritical regime. Beyond that energy, $p(U)$ gradually decreases and reaches zero at some critical value, say $U=U_c$. Once $U_c$ is crossed, all particle achieves sufficient kinetic energy to visit the full circle, clustering breaks down and the system reaches a paramagnetic state.
\par 

\begin{figure*}[ht!]
\centering
\includegraphics[width=5.75cm]{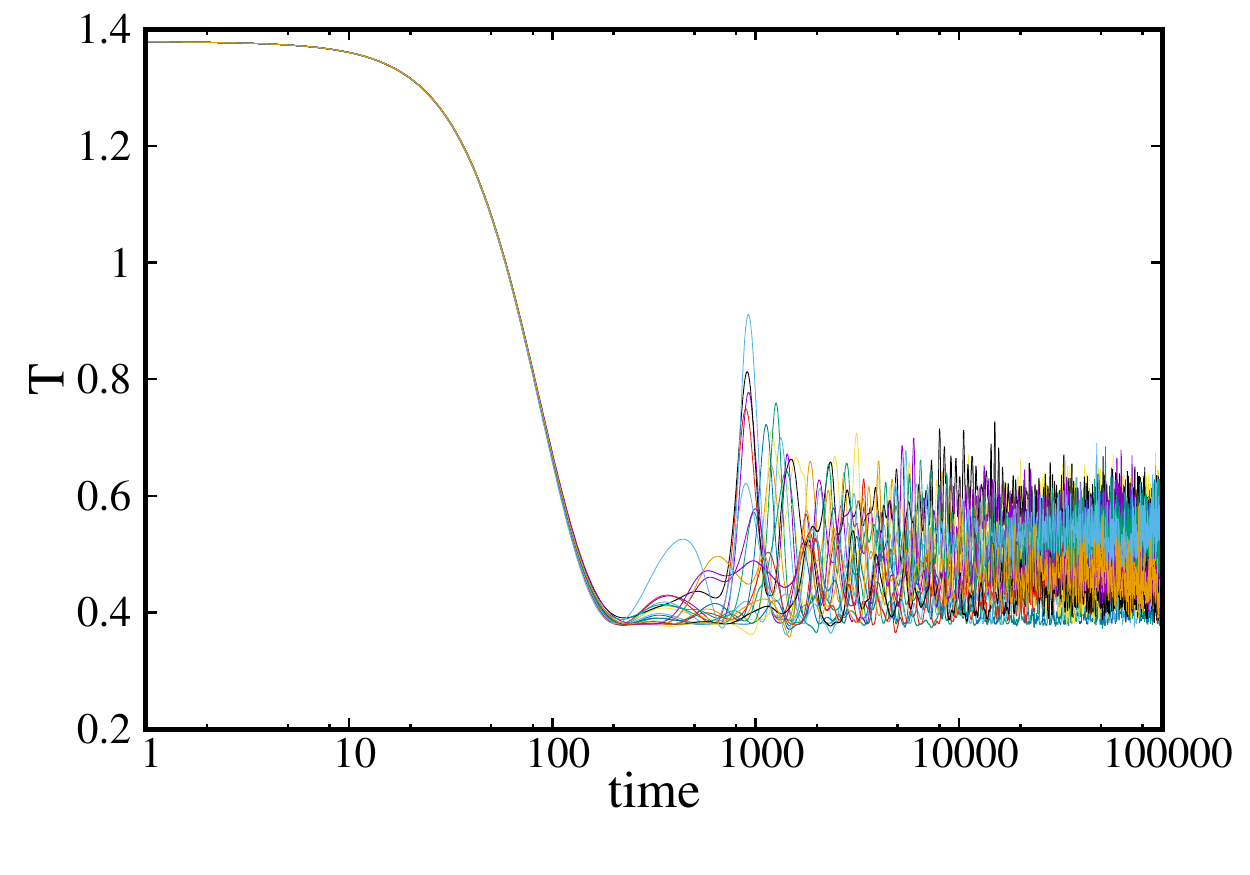}(a)\hspace{.05cm}
\includegraphics[width=5.75cm]{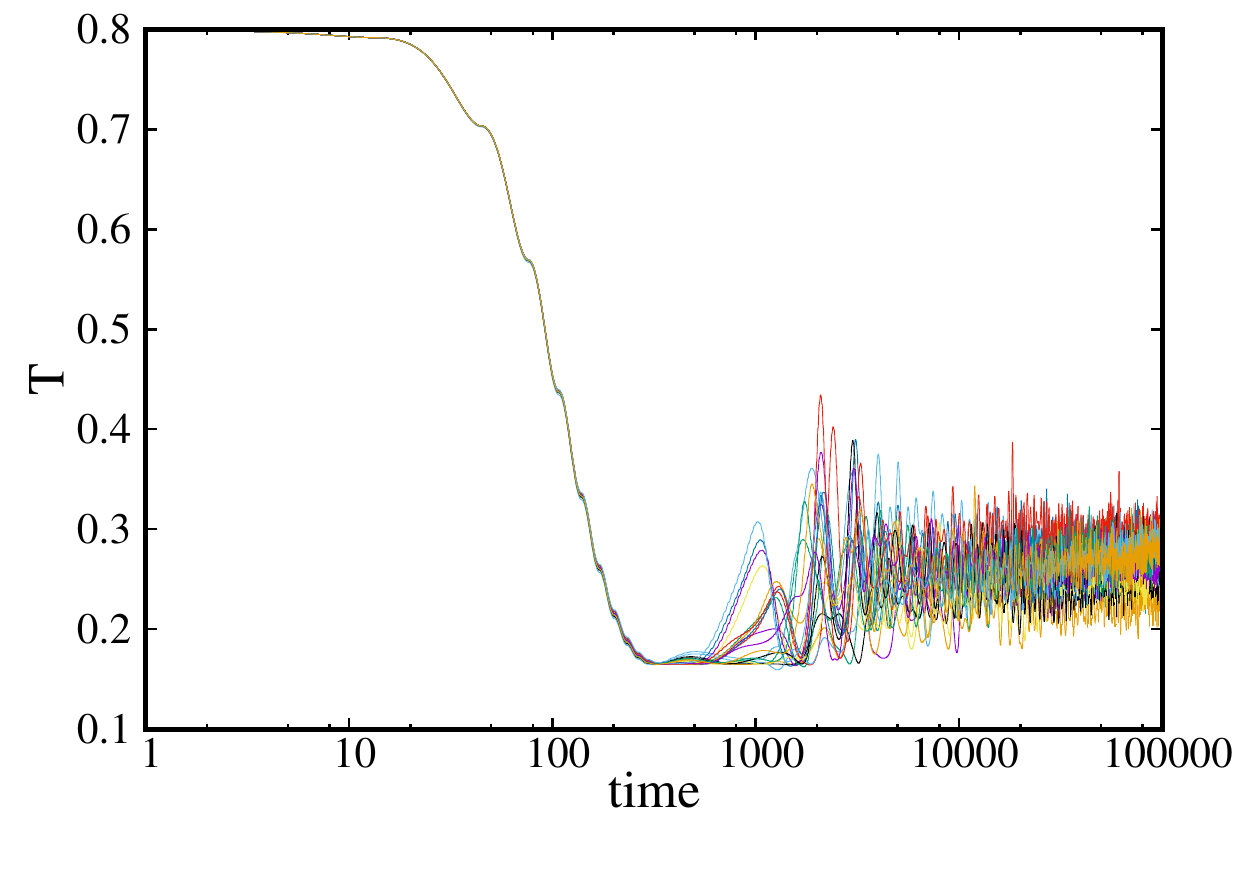}(b)
\vspace*{8pt}
\caption{(a) Time evolution of $T=2 \langle K \rangle/N$ for different initial conditions for $\omega=0$ ($20$ initial conditions shown in Fig.). Initial energy is $U=0.69$. Parameters are: $N=500, a=1.0$. After initial quick cooling the system reaches a plateau for a long time and then reaches its equilibrium temperature. (b) Time evolution of $T=2 \langle K \rangle/N$ for the energy density $U=0.4$ with different initial conditions (we have shown 20 in this figure) when $\omega=10$. Parameters are : $N=500, a=1.0$. After initial quick cooling the system reaches a plateau for a long time and then reaches its equilibrium temperature.}
\label{fig:QSSN500T}
\end{figure*}
\begin{figure}
\centering
\includegraphics[width=6cm]{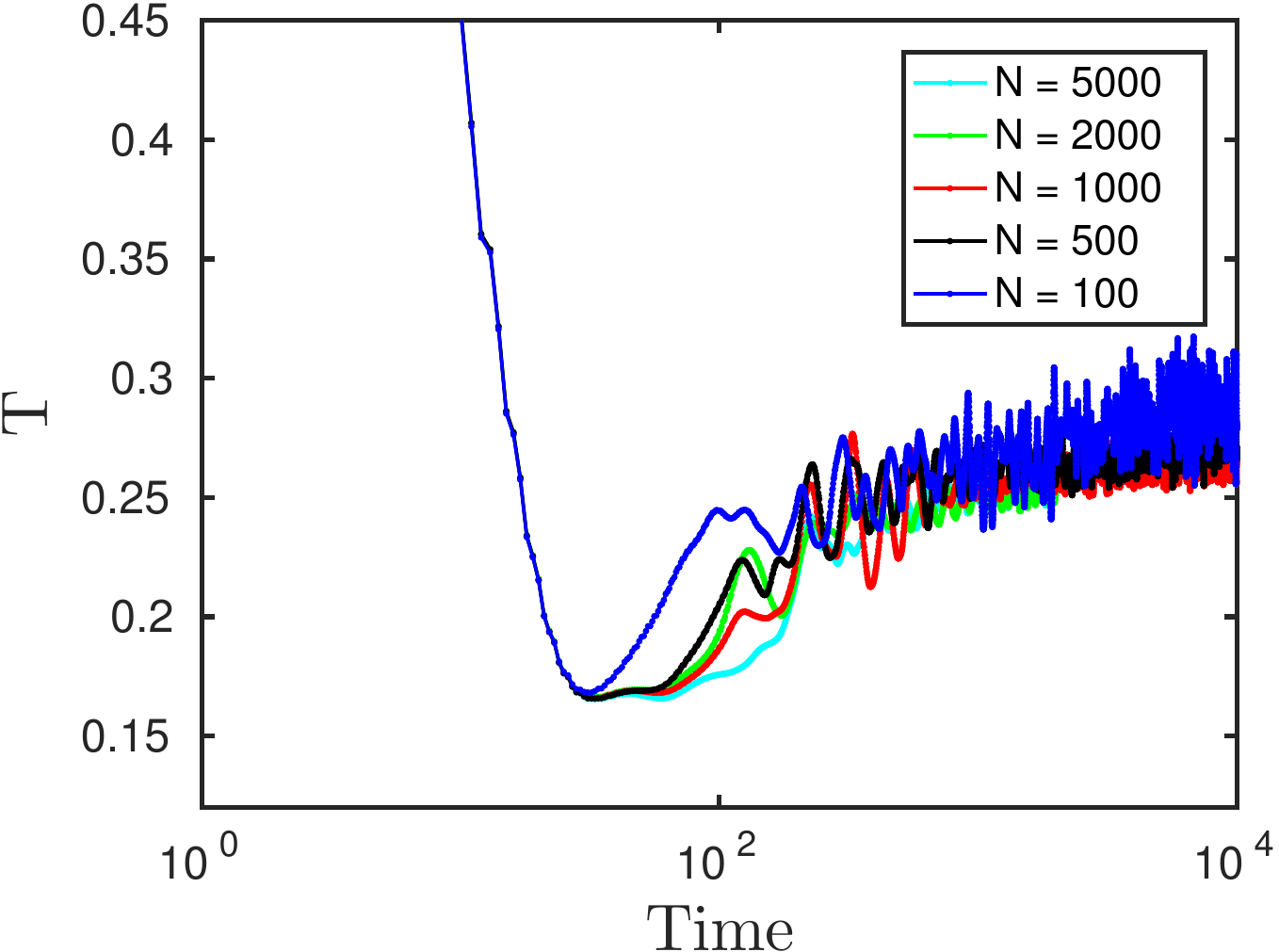}
\caption{ (Color online) Time evolution of temperature $T$ for different initial condition when $\omega=10$. Symbol represents different system sizes. Each point represents averaging over $1000$ different initial conditions. Initial energy is $U=0.4$ (before critical point $U_c\approx 0.49$) and initial magnetization $M_0=1$.}
\label{fig:systemsizerelation}
\end{figure}
Now we turn to the nonequilibrium properties of this model. One of the intriguing features of systems with long range interaction is the presence of quasistationary states - a state which is dynamically created and the lifetime depends on the system size $N$ \cite{Latora2001,Rapisarda2005,Pluchino2005,Yamaguchi1996,Pluchino2004,Baldovin2006,Pluchino2006}. The characeteristics of the QSS states depend on the initial configuration of the system \cite{Campa2007}. Microcanonical molecular dynamics simulations have shown that just below the energy density of the critical point ($U_c$) of the phase transition, the system reaches a QSS state, lifetime of which diverges with system size $N$. Therefore, it indicates that
if thermodynamic limit is considered, the system cannot relax towards Boltzmann-Gibbs equilibrium and will remain trapped in the QSS state. Noticeably, in the QSS state, the velocity distribution of the system is not Maxwellian \cite{Latora2001}. This phenomenolgy is widely observed in systems with long range interaction such as galaxy dynamics \cite{Padmanabh1990}, free electron lasers \cite{Barre2004} and 2D electron plasmas \cite{Kawahara2007}. This property of non-Gaussian velocity distribution has been interpreted in the framework of the statistical theory of the Vlasov equation which was first introduced in the context of astrophysical and 2D Euler turbulence \cite{Lynden-Bell1967,Chavanis2006}. It has been analytically proved that the QSS can be understood within a dynamical approach based on Vlasov equation \cite{Barre2006}. In fact, in the $N\rightarrow \infty$ limit, the one particle distribution of a specific class follows this model, which HMF model belongs to. The Vlasov equation governing the N-particle dynamics is given by
\begin{eqnarray}
\label{eq:vlasoveq}
\frac{\partial f}{\partial t}+p\frac{\partial f}{\partial \phi}-\frac{\partial V}{\partial \phi} \frac{\partial V}{\partial p}=0,
\end{eqnarray}
where $f(\phi, p, t)$ is the microscopic one-particle distribution. The stationary solution of this equation was understood on the theoretical framework of Lynden-Bell's analytical approach \cite{Lynden-Bell1967}. We expect that this approach is applicable for the modulated case as well. For $\omega=0$ and $a=1$, the system reduces to HMF model with uniform constant coupling parameter. Hence, Lyndell-Bell approach can be adopted in the modulated case as well. For $\omega\neq 0$, $V$ is modulated in Eq.\ref{eq:vlasoveq} by a temporal perturbation. However, the velocity profile we observe is a time averaged value over a certain time-span. Hence, we expect similar behaviour (modified by a constant factor) in the velocity profile even when $\omega\neq 0$. \par
The energy density mostly considered
in the literature is $0.69$, at equilibrium $M \approx 0.31$ which corresponds to a temperature
$T \approx 0.475$. Generally the initial conditions considered are those which correspond to
either $m = 1$ or $m = 0$. However, QSS states with different kind of initial conditions have
also been investigated\cite{Campa2007}. To realize this QSS state, ``\textit{water bag}" we consider inital conditions which consists of particles uniformly distributed in a rectangle $[-\phi_0,\phi_0] \times[-p_0,p_0]$ in the $(\phi,p)$ plane. The associated magnetization and specific energy are $M=M_0=\sin{\phi_0}/\phi_0$ and $U=\frac{p_0^2}{6}+\frac{1-M_0^2}{2}$. Once started with this set of initial conditions, the system gets frozen in the QSS state as discussed above. We simulate Eq. \ref{eq:HMFmodel_modulated} and show the time evolution of  $T=\frac{2\langle K \rangle}{N}$ in Fig. \ref{fig:QSSN500T}. For $\omega = 0 $, the temperature plateaux are shown for $U=0.69, N=1000$ in Fig. \ref{fig:QSSN500T}(a) (energy value where the anomaly is reported to be more evident). To understand the effect of the temporal modulation on this anomalous behaviour, we performed the simulation for $\omega=10$ and the energy is kept just below $U_c$ which is 0.4 (for this case, $U_c\approx 0.49$). Fig. \ref{fig:QSSN500T}(a) shows that the QSS appears in the modulated case as well and reaches the equilibrium value after certain relaxation period. To observe the  effect of system size we vary $N=100$ to $N=5000$ and show the time evolution of $T$(in log scale) in Fig.\ref{fig:systemsizerelation}. \par
\begin{figure*}[ht!]
\centering
\includegraphics[width=10cm]{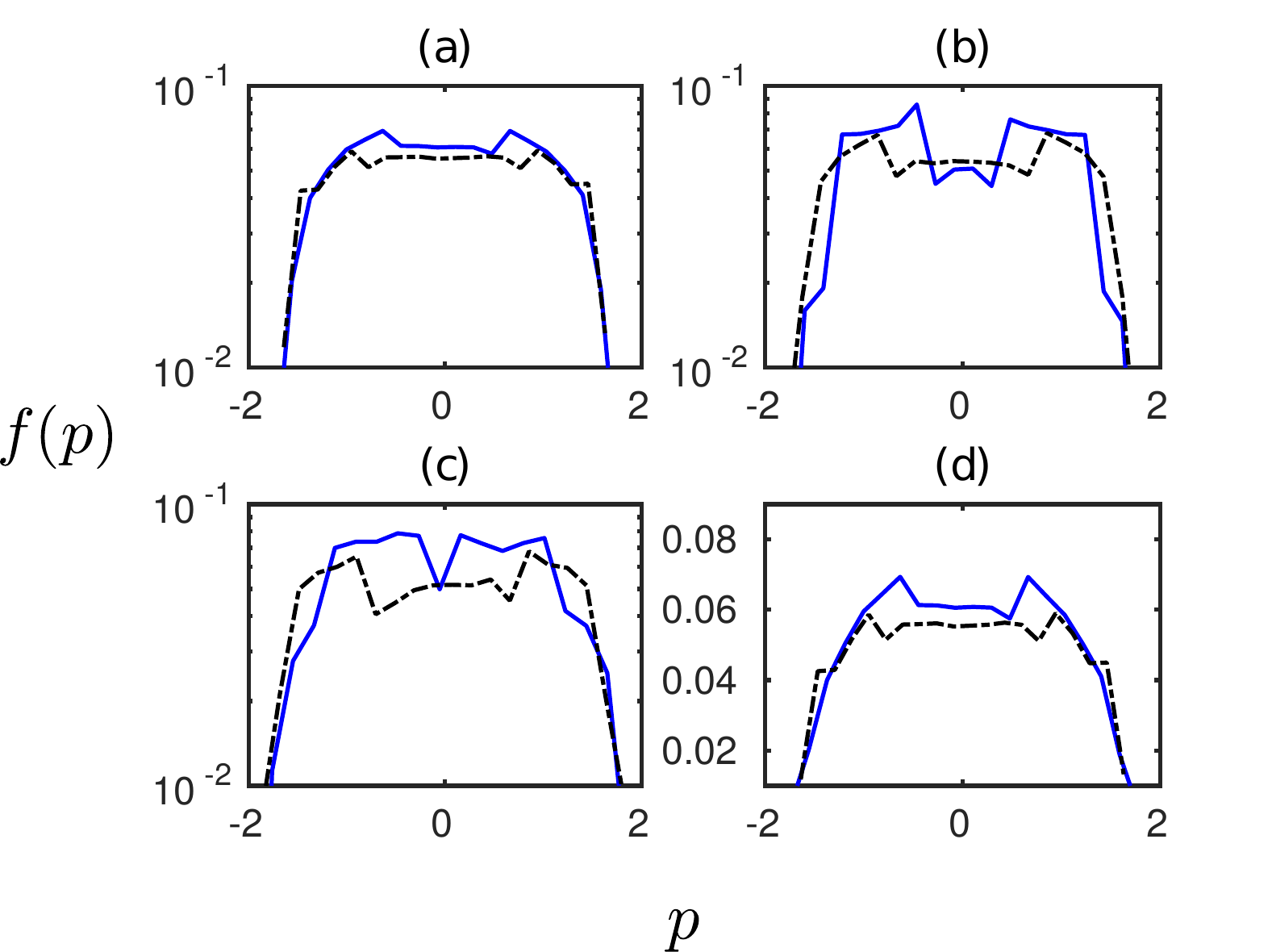}
\vspace*{8pt}
\caption{(Color online) Velocity distribution functions ($f(p)$).Solid lines stand for $f(p)$ when $\omega=0$ whereas, dashed lines stand for $f(p)$ when $\omega=10$. Panels (a), (b), (c) represents $f(p)$ for initial magnetization $M_0=0.3, 0.5$ and $0.7$ respectively in log-linear scale. Panel (d) shows $M=0.3$ in linear scale. The curves are computed for single realization with $N=5\times10^4$ in the time range $90<t<100$.
\label{fig:velocitydist} }
\end{figure*}
 For the constant coupling case, two symmetric bumps appear in the velocity distribution which is not a transient feature but a collective phenomenon\cite{Martelloni2016}. The formation of cluster of particles with opposite constant velocities starts early and the situation prevails during the following time. However, if initial magnetization
is increased, two bumps disappear and tend to merge to a single bump. A simple
dynamical argument has been provided in Ref.\cite{Antoniazzi2007a}. One-particle Hamiltonian can be written as $U=\frac{p^2}{2}-\epsilon( M_x\cos{\phi}+M_y\sin{\phi})$. For short times, $\phi\approx \phi_0+pt$. Eventually $M_x\approx(\sin{\delta _\phi}\sin{\delta_p}t)$ and $M_y\approx 0$. Hence, the energy becomes
\begin{eqnarray}
U(\phi,p)=\frac{p^2}{2}+\frac{\epsilon\sin{\delta_\phi}}{2\delta_\phi \delta_p t}\Big(\sin(\phi-\delta_p t)+\sin(\phi+\delta_p t)\Big),
\end{eqnarray}
which implies one particle interacting with two waves of phase velocities $\pm \delta p$. Depending on the initial condition, the particle is trapped in one of the two resonances. In Fig.\ref{fig:velocitydist}, we have shown the velocity distribution for both $\omega=0$ and $\omega=10$. We consider, $a=1, N=50000, U=0.88$. We observe presence of the symmetric bumps in the velocity distribution for the unperturbed case and unperturbed case as well.

\section{Discussion}
\label{sec:concl}

In this paper, we explored the effect of temporal modulation of the coupling parameter in HMF model. The equation (Eq.\ref{eq:modulatedHMFEOM}) governing this model essentially represents the equation of motion for a fully coupled driven pendula. The similar mathematical representation is shared by driven Bose Hubbard Hamiltonian (BHH) model where the potential is temporally modulated \cite{Salerno2014}. Experimental realization of this kind of modulation in Bose Hubbard model is possible with the aid of optical lattice set-up. Temporal modulation to the potential could be introduced by varying the intensity of the counter propagating lasers
forming the optical lattice. In several previous studies, this kind of temporal modulation in BHH model has been discussed in different contexts\cite{Witthaut2008,Boukobza2010}. In such a set-up, amplitude and frequency of modulation can be controlled precisely. Motivated by these studies, we introduced the temporal modulation to the HMF model, studied the effect in its equilibrium and prior to equilibrium state and compared with the constant coupling case. \par 
In order to study the system, N-body numerical simulations were carried out. We focus on the effect of the temporal modulation in the equilibrium phase transition. We observe a shift in the critical point from the constant coupling case. The shift in $U_c$ is independent of the modulation frequency for $\omega>2$. The effect of amplitude of modulation $a$ is, however, dominant. We also show the finite size effect on the shift of the critical point in this context. From the trapping probability we found that for the modulated case the particles achieve kinetic energy enough to visit the full circle much before the $U_c$ for the constant coupling case.  Hence, clustered phase is lost and the system reaches a homogeneous phase. \par 
Prior to its equilibrium, our numerical results show that the QSS persists even in the presence of the proposed modulation for the HMF model. A symmetric bump is observed in the velocity distribution for the temporally modulated case as well. For the constant coupling case, previous studies have discussed a phase transition between homogeneous and inhomogenous QSS \cite{Chavanis2005,Antoniazzi2007a,Antoniazzi2007b}. At initial magnetization $M_0=0.897$, for $U=0.69$ a bifurcation appears. Below this critical value of initial magnetization, the maximum entropy principle predics a zero magnetization($M_{\text{QSS}}$) value. However, above this point, $M_{\text{QSS}}$ develops a nonzero value. Numerical analysis can be performed to observe the effect of temporal modulation in this out-of-equilibrium phase transition in the system.

\section{Acknowledgements}
The authors would like to thank  Anandamohan Ghosh, Gopal Sardar and Biswarup Ash for valuable discussions and suggestions.



\begin{thebibliography}{10}

\bibitem{Antoni1995}
Mickael Antoni and Stefano Ruffo.
\newblock Clustering and relaxation in hamiltonian long-range dynamics.
\newblock {\em Phys. Rev. E}, 52:2361--2374, Sep 1995.

\bibitem{Campa2009}
Alessandro Campa, Thierry Dauxois, and Stefano Ruffo.
\newblock Statistical mechanics and dynamics of solvable models with long-range
  interactions.
\newblock {\em Physics Reports}, 480(3):57 -- 159, 2009.

\bibitem{Campa2014}
Alessandro Campa.
\newblock {\em Physics of long-range interacting systems}.
\newblock Oxford University Press, Oxford, 2014.

\bibitem{Levin2014}
Yan Levin, Renato Pakter, Felipe~B. Rizzato, Tarcísio~N. Teles, and
  Fernanda~P.C. Benetti.
\newblock Nonequilibrium statistical mechanics of systems with long-range
  interactions.
\newblock {\em Physics Reports}, 535(1):1 -- 60, 2014.
\newblock Nonequilibrium statistical mechanics of systems with long-range
  interactions.

\bibitem{Latora2001}
Vito Latora, Andrea Rapisarda, and Constantino Tsallis.
\newblock Non-gaussian equilibrium in a long-range hamiltonian system.
\newblock {\em Phys. Rev. E}, 64:056134, Oct 2001.

\bibitem{Antoniazzi2006}
A.~Antoniazzi, Y.~Elskens, D.~Fanelli, and S.~Ruffo.
\newblock Statistical mechanics and vlasov equation allow for a
  simplifiedhamiltonian description of single-pass free electron lasersaturated
  dynamics.
\newblock {\em The European Physical Journal B - Condensed Matter and Complex
  Systems}, 50(4):603--611, Apr 2006.

\bibitem{Elskens2003}
Y~Elskens and D~Escande.
\newblock Microscopic dynamics of plasmas and chaos.
\newblock {\em Plasma Physics and Controlled Fusion}, 45(4):521, 2003.

\bibitem{Benedetti2006}
C~Benedetti, S~Rambaldi, and G~Turchetti.
\newblock Relaxation to boltzmann eqpisardauilibrium of 2d coulomb oscillators.
\newblock {\em Physica A: Statistical Mechanics and its Applications},
  364:197--212, 2006.

\bibitem{Lynden-Bell1967}
D.~{Lynden-Bell}.
\newblock {Statistical mechanics of violent relaxation in stellar systems}.
\newblock { }, 136:101, 1967.

\bibitem{Montemurro2003}
Marcelo~A Montemurro, Francisco~A Tamarit, and Celia Anteneodo.
\newblock Aging in an infinite-range hamiltonian system of coupled rotators.
\newblock {\em Physical Review E}, 67(3):031106, 2003.

\bibitem{Pluchino2004a}
Alessandro Pluchino, Vito Latora, and Andrea Rapisarda.
\newblock Glassy dynamics in the hmf model.
\newblock {\em Physica A: Statistical Mechanics and its Applications},
  340(1):187 -- 195, 2004.
\newblock News and Expectations in Thermostatistics.

\bibitem{Rapisarda2005}
Andrea Rapisarda and Alessandro Pluchino.
\newblock Nonextensive thermodynamics and glassy behaviour.
\newblock {\em Europhysics News}, 36(6):202--206, 2005.

\bibitem{Pluchino2005}
A.~{Pluchino}, A.~{Rapisarda}, and V.~{Latora}.
\newblock {Metastability and Anomalous Behavior in the Hmf Model:. Connections
  to Nonextensive Thermodynamics and Glassy Dynamics}.
\newblock In C.~{Beck}, G.~{Benedek}, A.~{Rapisarda}, and C.~{Tsallis},
  editors, {\em Complexity, Metastability and Nonextensivity}, pages 102--112,
  September 2005.

\bibitem{Pluchino2006}
Alessandro Pluchino and Andrea Rapisarda.
\newblock Metastability in the hamiltonian mean field model and kuramoto model.
\newblock {\em Physica A: Statistical Mechanics and its Applications},
  365(1):184 -- 189, 2006.
\newblock Fundamental Problems of Modern Statistical MechanicsProceedings of
  the 3rd International Conference on 'News, Expectations and Trends in
  Statistical Physics'News, Expectations and Trends in Statistical Physics.

\bibitem{Tauro2007}
C.~B. Tauro, G.~Maglione, and F.~A. Tamarit.
\newblock Relaxation dynamics and topology in the hamiltonian mean field model.
\newblock {\em The European Physical Journal Special Topics}, 143(1):9--12, Apr
  2007.

\bibitem{Antoniazzi2007}
Andrea Antoniazzi, Francesco Califano, Duccio Fanelli, and Stefano Ruffo.
\newblock Exploring the thermodynamic limit of hamiltonian models: Convergence
  to the vlasov equation.
\newblock {\em Physical review letters}, 98(15):150602, 2007.

\bibitem{Leoncini2009}
X.~Leoncini, T.~L. Van~Den Berg, and D.~Fanelli.
\newblock Out-of-equilibrium solutions in the xy-hamiltonian mean-field model.
\newblock {\em EPL (Europhysics Letters)}, 86(2):20002, 2009.

\bibitem{Barre2006}
Julien Barré, Freddy Bouchet, Thierry Dauxois, Stefano Ruffo, and Yoshiyuki~Y.
  Yamaguchi.
\newblock The vlasov equation and the hamiltonian mean-field model.
\newblock {\em Physica A: Statistical Mechanics and its Applications},
  365(1):177 -- 183, 2006.
\newblock Fundamental Problems of Modern Statistical MechanicsProceedings of
  the 3rd International Conference on 'News, Expectations and Trends in
  Statistical Physics'News, Expectations and Trends in Statistical Physics.

\bibitem{Campa2007}
Alessandro Campa, Andrea Giansanti, and Gianluca Morelli.
\newblock Long-time behavior of quasistationary states of the hamiltonian
  mean-field model.
\newblock {\em Phys. Rev. E}, 76:041117, Oct 2007.

\bibitem{Konishi1992}
T~Konishi and K~Kaneko.
\newblock Clustered motion in symplectic coupled map systems.
\newblock {\em Journal of Physics A: Mathematical and General}, 25(23):6283,
  1992.

\bibitem{Medvedyeva2003}
Kateryna Medvedyeva, Petter Holme, Petter Minnhagen, and Beom~Jun Kim.
\newblock Dynamic critical behavior of the $\mathrm{XY}$ model in small-world
  networks.
\newblock {\em Phys. Rev. E}, 67:036118, Mar 2003.

\bibitem{Kim2001}
Beom~Jun Kim, H.~Hong, Petter Holme, Gun~Sang Jeon, Petter Minnhagen, and M.~Y.
  Choi.
\newblock Xy.
\newblock {\em Phys. Rev. E}, 64:056135, Oct 2001.

\bibitem{DeNigris2013}
Sarah De~Nigris and Xavier Leoncini.
\newblock Critical behavior of the $xy$-rotor model on regular and small-world
  networks.
\newblock {\em Phys. Rev. E}, 88:012131, Jul 2013.

\bibitem{Restrepo2014}
Juan~G. Restrepo and James~D. Meiss.
\newblock Onset of synchronization in the disordered hamiltonian mean-field
  model.
\newblock {\em Phys. Rev. E}, 89:052125, May 2014.

\bibitem{Virkar2015}
Yogesh~S. Virkar, Juan~G. Restrepo, and James~D. Meiss.
\newblock Hamiltonian mean field model: Effect of network structure on
  synchronization dynamics.
\newblock {\em Phys. Rev. E}, 92:052802, Nov 2015.

\bibitem{Ciani2010}
Antonia Ciani, Duccio Fanelli, and Stefano Ruffo.
\newblock {\em Long-range Interactions and Diluted Networks}, pages 83--132.
\newblock Springer Berlin Heidelberg, Berlin, Heidelberg, 2010.

\bibitem{Watts1998}
Duncan~J. Watts and Steven~H. Strogatz.
\newblock Collective dynamics of /`small-world/' networks.
\newblock {\em Nature}, 393(6684):440--442, Jun 1998.

\bibitem{Luo2011}
Albert Luo and Valentin Afraimovich.
\newblock {\em Long-range Interactions, Stochasticity and Fractional Dynamics:
  Dedicated to George M. Zaslavsky (1935—2008)}.
\newblock Springer Science \& Business Media, 2011.

\bibitem{Nigris2013}
Sarah~De Nigris and Xavier Leoncini.
\newblock Emergence of a non-trivial fluctuating phase in the xy-rotors model
  on regular networks.
\newblock {\em EPL (Europhysics Letters)}, 101(1):10002, 2013.

\bibitem{Stojanovski1997}
Toni Stojanovski, Ljupco Kocarev, Ulrich Parlitz, and Richard Harris.
\newblock Sporadic driving of dynamical systems.
\newblock {\em Physical Review E}, 55(4):4035, 1997.

\bibitem{Ito2001}
Junji Ito and Kunihiko Kaneko.
\newblock Spontaneous structure formation in a network of chaotic units with
  variable connection strengths.
\newblock {\em Phys. Rev. Lett.}, 88:028701, Dec 2001.

\bibitem{Zanette2004}
Dami{\'a}n~H Zanette and Alexander~S Mikhailov.
\newblock Dynamical systems with time-dependent coupling: clustering and
  critical behaviour.
\newblock {\em Physica D: Nonlinear Phenomena}, 194(3):203--218, 2004.

\bibitem{Masuda2016}
Naoki Masuda.
\newblock {\em A guide to temporal networks}.
\newblock World Scientific, New Jersey, 2016.

\bibitem{Petit2017}
Julien Petit, Ben Lauwens, Duccio Fanelli, and Timoteo Carletti.
\newblock The theory of turing patterns on time varying networks.
\newblock {\em arXiv preprint arXiv:1705.08025}, 2017.

\bibitem{Stilwell2006}
Daniel~J. Stilwell, Erik~M. Bollt, and D.~Gray Roberson.
\newblock Sufficient conditions for fast switching synchronization in
  time-varying network topologies.
\newblock {\em SIAM Journal on Applied Dynamical Systems}, 5(1):140--156, 2006.

\bibitem{konishi100}
{T. Konishi and K. Kaneko, J. Phys. A {25}, 6283 (1992); M. Antoni and S.
  Ruffo, Phys. Rev. E {52}, 2361 (1995)}.

\bibitem{Escande1994}
Dominique Escande, Holger Kantz, Roberto Livi, and Stefano Ruffo.
\newblock Self-consistent check of the validity of gibbs calculus using
  dynamical variables.
\newblock {\em Journal of Statistical Physics}, 76(1):605--626, 1994.

\bibitem{Yamaguchi1996}
Yoshiyuki~Y. Yamaguchi.
\newblock Slow relaxation at critical point of second order phase transition in
  a highly chaotic hamiltonian system.
\newblock {\em Progress of Theoretical Physics}, 95(4):717--731, 1996.

\bibitem{Pluchino2004}
Alessandro Pluchino, Vito Latora, and Andrea Rapisarda.
\newblock Dynamical anomalies and the role of initial conditions in the \{HMF\}
  model.
\newblock {\em Physica A: Statistical Mechanics and its Applications},
  338(1–2):60 -- 67, 2004.
\newblock Proceedings of the conference A Nonlinear World: the Real World, 2nd
  International Conference on Frontier Science.

\bibitem{Baldovin2006}
Fulvio Baldovin and Enzo Orlandini.
\newblock Hamiltonian dynamics reveals the existence of quasistationary states
  for long-range systems in contact with a reservoir.
\newblock {\em Phys. Rev. Lett.}, 96:240602, Jun 2006.

\bibitem{Padmanabh1990}
T.~Padmanabhan.
\newblock Statistical mechanics of gravitating systems.
\newblock {\em Physics Reports}, 188(5):285 -- 362, 1990.

\bibitem{Barre2004}
Julien Barr\'e, Thierry Dauxois, Giovanni De~Ninno, Duccio Fanelli, and Stefano
  Ruffo.
\newblock Statistical theory of high-gain free-electron laser saturation.
\newblock {\em Phys. Rev. E}, 69:045501, Apr 2004.

\bibitem{Kawahara2007}
Ryo Kawahara and Hiizu Nakanishi.
\newblock Slow relaxation in two-dimensional electron plasma under strong
  magnetic field.
\newblock {\em Journal of the Physical Society of Japan}, 76(7):074001, 2007.

\bibitem{Chavanis2006}
Pierre-Henri Chavanis.
\newblock Quasi-stationary states and incomplete violent relaxation in systems
  with long-range interactions.
\newblock {\em Physica A: Statistical Mechanics and its Applications},
  365(1):102 -- 107, 2006.
\newblock Fundamental Problems of Modern Statistical Mechanics.

\bibitem{Martelloni2016}
Gabriele Martelloni, Gianluca Martelloni, Pierre de~Buyl, and Duccio Fanelli.
\newblock Generalized maximum entropy approach to quasistationary states in
  long-range systems.
\newblock {\em Phys. Rev. E}, 93:022107, Feb 2016.

\bibitem{Antoniazzi2007a}
Andrea Antoniazzi, Duccio Fanelli, Julien Barr\'e, Pierre-Henri Chavanis,
  Thierry Dauxois, and Stefano Ruffo.
\newblock Maximum entropy principle explains quasistationary states in systems
  with long-range interactions: The example of the hamiltonian mean-field
  model.
\newblock {\em Phys. Rev. E}, 75:011112, Jan 2007.

\bibitem{Salerno2014}
Grazia Salerno and Iacopo Carusotto.
\newblock Dynamical decoupling and dynamical isolation in temporally modulated
  coupled pendulums.
\newblock {\em EPL (Europhysics Letters)}, 106(2):24002, 2014.

\bibitem{Witthaut2008}
D.~Witthaut, F.~Trimborn, and S.~Wimberger.
\newblock Dissipation induced coherence of a two-mode bose-einstein condensate.
\newblock {\em Phys. Rev. Lett.}, 101:200402, Nov 2008.

\bibitem{Boukobza2010}
Erez Boukobza, Michael~G. Moore, Doron Cohen, and Amichay Vardi.
\newblock Nonlinear phase dynamics in a driven bosonic josephson junction.
\newblock {\em Phys. Rev. Lett.}, 104:240402, Jun 2010.

\bibitem{Chavanis2005}
{Chavanis, P. H.}, {Vatteville, J.}, and {Bouchet, F.}
\newblock Dynamics and thermodynamics of a simple model similar to
  self-gravitating systems: the hmf model.
\newblock {\em Eur. Phys. J. B}, 46(1):61--99, 2005.

\bibitem{Antoniazzi2007b}
Andrea Antoniazzi, Duccio Fanelli, Stefano Ruffo, and Yoshiyuki~Y. Yamaguchi.
\newblock Nonequilibrium tricritical point in a system with long-range
  interactions.
\newblock {\em Phys. Rev. Lett.}, 99:040601, Jul 2007.

\end{thebibliography}

\end{document}